\documentclass[aps,prmaterials,reprint,citeautoscript,superscriptaddress]{revtex4-1}

\usepackage{amsmath}
\usepackage{amsfonts}
\usepackage{amssymb}
\usepackage{graphicx}
\usepackage{color}
\usepackage{subfigure}

\newcommand{\algao}{(Al$_x$Ga$_{1{\rm -}x})_2$O$_3$}
\newcommand{\ingao}{(In$_x$Ga$_{1{\rm -}x})_2$O$_3$}
\newcommand{\inalo}{(In$_x$Al$_{1{\rm -}x})_2$O$_3$}
\begin{document}

\title{Computational design of optimal heterostructures for $\beta$-Ga$_2$O$_3$}

\author{Sierra Seacat}
\affiliation{Department of Physics and Astronomy, University of Kansas, Lawrence, KS 66045, USA}
\email{scseacat@ku.edu}

\author{John L. Lyons}
\affiliation{Center for Computational Materials Science, US Naval Research Laboratory, Washington, DC 20375, USA}

\author{Hartwin Peelaers}
\affiliation{Department of Physics and Astronomy, University of Kansas, Lawrence, KS 66045, USA}
\email{peelaers@ku.edu}

\begin{abstract}
Ga$_2$O$_3$ is a wide-bandgap material of interest for a wide variety of devices, many of these requiring heterostructures, for instance to achieve carrier confinement. A common method to create such heterostructures is to alloy with In$_2$O$_3$ or Al$_2$O$_3$. However, the lattice constants of these materials are significantly different from those of Ga$_2$O$_3$, leading to large amounts of strain in the resulting heterostructure. If the thickness of the heterostructure is increased, this can lead to cracking. By considering alloys of In$_2$O$_3$ and Al$_2$O$_3$, the lattice constants can be tailored to those of Ga$_2$O$_3$, while still keeping a sizable conduction-band offset. We use density functional theory with hybrid functionals to investigate the structural and electronic properties of In$_2$O$_3$ and Al$_2$O$_3$ alloys in the bixbyite, corundum, and monoclinic structures. We find that the lattice constants increase with In incorporation. Bandgaps decrease nonlinearly with increasing In concentration. We find the (In$_{\rm 0.25}$Al$_{\rm 0.75}$)$_{\rm 2}$O$_{\rm 3}$ 
monoclinic structure to be of particular interest, as it closely matches the Ga$_2$O$_3$ lattice constants while providing an indirect/direct bandgap of 5.94/5.70 eV and a conduction-band offset of 1 eV compared to Ga$_2$O$_3$.

\end{abstract}

\maketitle

As a wide-bandgap semiconductor (4.7 eV)~\cite{Tippins1965,Matsumoto1974}, Ga$_2$O$_3$ is a promising material for solar blind photodetectors, which have a variety of applications across different fields including military surveillance, medical imaging, chemical and biological analysis, flame detection, and secure communications~\cite{Suzuki2009,Hou2021, Guo2019, Zou2014, Zhang2020}. Additionally, Ga$_2$O$_3$ possesses a large breakdown field (5-9 MVcm$^{-1}$)~\cite{Pearton2018b,Higashiwaki2017}, which makes it a suitable material for high-power devices such as field effect transistors and Schottky diodes~\cite{Higashiwaki2017, Pearton2018a}.

Fabricating such devices commonly requires creating heterostructures with a conduction-band offset to confine charge carriers. 
One way to achieve this without changing the crystal structure is by alloying Ga$_2$O$_3$ with Al$_2$O$_3$~\cite{Ahmadi2017a,Krishnamoorthy2017,Zhang2018d,Joishi2019,Okumura2019,Chatterjee2020,Vaidya2021,Tadjer2021}. 

\begin{figure}[tb]
	\includegraphics[width=0.92\columnwidth]{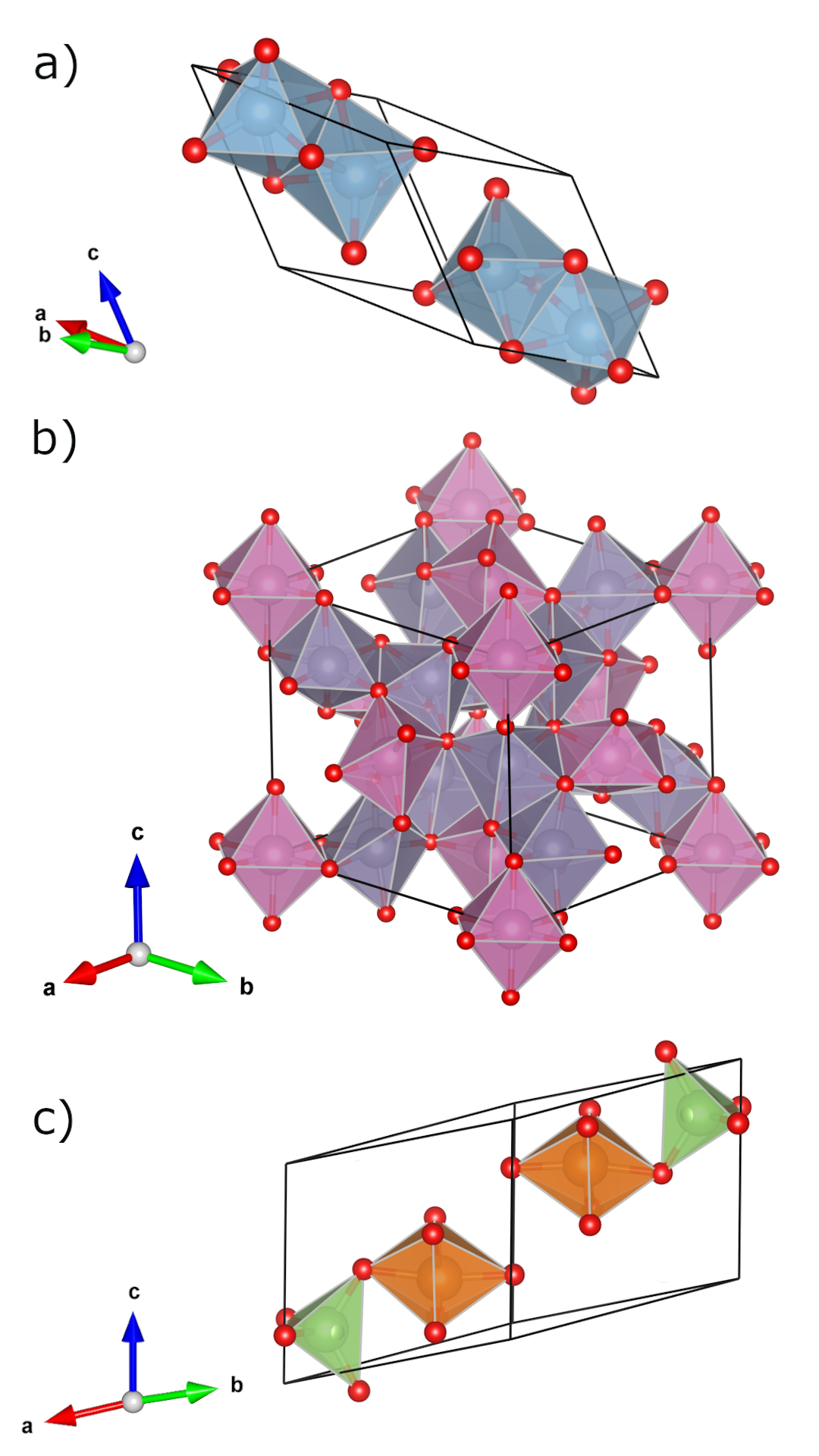}
	\caption{\label{fig:structures} a) The primitive unit cell of corundum Al$_2$O$_3$. b) The primitive unit cell of bixbyite In$_2$O$_3$ with the distorted octahedra shown in dark gray/purple and the regular octahedra sites shown in light pink/purple. c) The primitive unit cell of monoclinic Ga$_2$O$_3$ with the cations colored by site coordination. The octahedral sites are shown in orange, while the tetrahedral sites are shown in green. }
\end{figure}

In and Al, being in the same column in the periodic table as Ga, are common candidates for alloying with Ga$_2$O$_3$~\cite{Peelaers2018, Peelaers2015,Seacat2020,Seacat2021,Wang2018b,Kim2021b,Wouters2020}. However, the ground state structures of Ga$_2$O$_3$, In$_2$O$_3$, and Al$_2$O$_3$ are all different: Ga$_2$O$_3$ occurs in a monoclinic crystal structure [referred to as $\beta$-Ga$_2$O$_3$, and shown in Fig.~\ref{fig:structures}(a)], with C2/m symmetry, In$_2$O$_3$ in a cubic bixbyite structure with Ia$\bar{3}$ symmetry [Fig.~\ref{fig:structures}(b)], and Al$_2$O$_3$ in a corundum structure with R3c symmetry (known as $\alpha$-Al$_2$O$_3$) [Fig.~\ref{fig:structures}(c)]. The lattice constants of these three materials are all different, with those of In$_2$O$_3$ being the largest, followed by Ga$_2$O$_3$, and those of Al$_2$O$_3$ being the smallest. This ordering of lattice constant lengths occurs for both the ground state structures and for the (meta)stable polymorphs~\cite{Peelaers2018, Peelaers2015,Seacat2020,Seacat2021}.

This mismatch in lattice constants is undesirable, as it creates strain in the resulting alloy. If the thickness of the heterostructure is sufficiently large, cracking can occur~\cite{Mu2020,Lundh2023,Ohtsuki2023}. 
Here we propose a solution to this strain by considering alloys of In$_2$O$_3$ with Al$_2$O$_3$, which we can tailor to lattice-match monoclinic Ga$_2$O$_3$, to create optimal heterostructures. 

We use density functional theory (DFT) at the hybrid functional level to investigate the structural and electronic properties of~\inalo~alloys. We determine the lowest energy structures for each In concentration, considering the ground state structures (bixbyite and corundum), as well as the monoclinic structure (the desired structure as it is the ground state structure of Ga$_2$O$_3$). The lowest energy configurations for each of the structures are subsequently used to find the changes in lattice constants, bandgaps, and valence- and conduction-band offsets as a function of In concentration. As expected from Vegard's law, the lattice constants increase linearly with In content, as In$_2$O$_3$ has larger lattice constants compared to Al$_2$O$_3$. In contrast, the bandgap decreases non-linearly with increasing In content, since the bandgap of In$_2$O$_3$ is 3.14 eV, and thus smaller than the 8.81 eV bandgap of Al$_2$O$_3$. We find that an alloy with 25\% In and 75\% Al forms a perfect lattice match with Ga$_2$O$_3$, while still having a large conduction-band offset. Considering In$_2$O$_3$-Al$_2$O$_3$ alloys therefore increases the design space for devices, allowing for different bandgap and lattice constant values than those that are possible when only looking at \algao~and \ingao~alloys.

All DFT calculations were performed using projector augmented wave (PAW) potentials~\cite{Blochl1994} as implemented in the Vienna ab initio simulation package (VASP) \cite{Kresse1993,Kresse1996}. In order to obtain the necessary accuracy for the structural and electronic properties, the Heyd-Scuseria-Ernzerhof (HSE06) hybrid functional with the mixing parameter set to 32\% was used~\cite{Heyd2003,*Heyd2006}. To ensure consistency with previous results and to guarantee accurate energetic ordering, the In pseudopotential with $d$ electrons as part of the valence was used~\cite{Peelaers2018,Mu2019,Sabino2014}.  All calculations were performed using a plane-wave expansion cutoff of 500 eV with all structures relaxed until the forces were smaller than 10 meV/\AA ~and all stresses were smaller than 0.5 meV/\AA$^{3}$. A Monkhorst-Pack 4$\times$4$\times$4 {\bf k}-point grid was used for the corundum and primitive monoclinic calculations, a 6$\times$4$\times$4 {\bf k}-point grid for the calculations using the conventional monoclinic cell, and a 2$\times$2$\times$2 {\bf k}-point grid for the bixbyite calculations. In order to align the band edges of the alloys on an absolute energy scale using the vacuum level, we also calculated oxide surfaces, which were generated from primitive alloy cells. Non-polar (110) surfaces are used, with at least 25~\AA~of oxide and 25~\AA~of vacuum.

\begin{table}[tb]
	\caption{\label{tab:vals} Calculated and experimental lattice constants and band gaps for the ground state structures of Ga$_2$O$_3$, Al$_2$O$_3$, and In$_2$O$_3$.}
	\begin{ruledtabular}
		\begin{center}
	\centering
	\begin{tabular}{lllllll}
		& \multicolumn{2}{l}{$\beta$-Ga$_2$O$_3$} & \multicolumn{2}{l}{$\alpha$-Al$_2$O$_3$} & \multicolumn{2}{l}{In$_2$O$_3$}  \\
 		& Calc. & Exp. & Calc. & Exp. & Calc. & Exp. \\
		\hline
		Symmetry	& \multicolumn{2}{l}{monoclinic} & \multicolumn{2}{l}{corundum} & \multicolumn{2}{l}{bixbyite} \\
		Spacegroup & \multicolumn{2}{l}{C2/m} & \multicolumn{2}{l}{R3c} & \multicolumn{2}{l}{Ia$\bar{3}$} \\
		$a$ (\AA)  & 12.25 & 12.21$^a$ & 4.74 & 4.76$^c$ & 10.11 & 10.12$^e$ \\
		$b$ (\AA)  & 3.04 & 3.04$^a$ & 4.74  & 4.76$^c$ &		& 		     \\
		$c$ (\AA)  & 5.79 & 5.80$^a$ & 12.94 & 12.99$^c$ &		& 	           \\
		$\beta$	& 103.82$^\circ$  & 103.83$^\circ$$^a$  &		&		  &         &   	\\
		$E^{\rm direct}_{\rm gap}$ (eV) & 4.83 & 4.76$^b$ & 8.81 & 8.8$^d$ & 3.14 & 2.9$^f$ \\
	\end{tabular}
\end{center}
\end{ruledtabular}
\raggedright
$^a$Ref.~\onlinecite{Aahman1996}\\
$^b$Ref.~\onlinecite{Matsumoto1974b}\\
$^c$Ref.~\onlinecite{Haan1962}\\
$^d$Ref.~\onlinecite{French1990}\\
$^e$Ref.~\onlinecite{Marezio1966}\\
$^f$Ref.~\onlinecite{Irmscher2014}
\end{table}

We consider alloys with monoclinic, corundum, and bixbyite structures. For the monoclinic alloys, both the 10-atom primitive unit cell and the 20-atom conventional unit cell are used to perform calculations. We use the 10-atom primitive unit cell for the corundum alloys, and bixbyite calculations are performed using the 40-atom primitive unit cell. Our calculated lattice constants and bandgaps are listed in Table~\ref{tab:vals} with a comparison to experimental values.

The different ground state structures, shown in Fig.~\ref{fig:structures}, have different cation bonding environments. The corundum and bixbyite cells only have octahedrally coordinated cations, while the primitive monoclinic cell consists of two octahedrally coordinated and two tetrahedrally coordinated cation sites. All the cation sites in the corundum cell are identical, while there are two different types of octahedral sites in the bixbyite structure: regular octahedra with longer bond lengths and distorted octahedra with shorter bond lengths. To determine the lowest energy structures for each composition, these differences in site coordination are considered.

We use the enthalpy of formation ($\Delta H$) to characterize the relative stability of each possible alloy, which is defined as:
\begin{align}\label{eq:enthalpy}
& \Delta H[(\text{In}_{x}\text{Al}_{1-x})_2\text{O}_3] = E[(\text{In}_{x}\text{Al}_{1-x})_2\text{O}_3] \notag \\
&\qquad- (1-x)\,E[\text{Al}_2\text{O}_3] - x\,E[\text{In}_2\text{O}_3] \ ,
\end{align}
where $E[\text{Al}_2\text{O}_3]$ is the energy per f.u.~of corundum Al$_2$O$_3$ and $E[\text{In}_2\text{O}_3]$ is the energy per f.u.~of bixbyite In$_2$O$_3$.

We first determine the most favorable cation coordination environments for the monoclinic and bixbyite cells. Starting with monoclinic Al$_2$O$_3$, we replace one Al atom with an In atom for both an octahedral and a tetrahedral site. The enthalpy of formation is then calculated for each structure to determine the most likely configuration. The results for a single In atom replacement are then used to repeat the analysis for larger concentrations of In.  A similar procedure is followed for the bixbyite structure, considering the distorted and undistorted octahedral sites.
In the monoclinic structure, In prefers to occupy octahedral sites. For the bixbyite structure, single Al or In atoms prefer to occupy the regular octahedra, but this preference is lost at higher concentrations.

\begin{figure}[tb]
	\includegraphics[width=0.95\columnwidth]{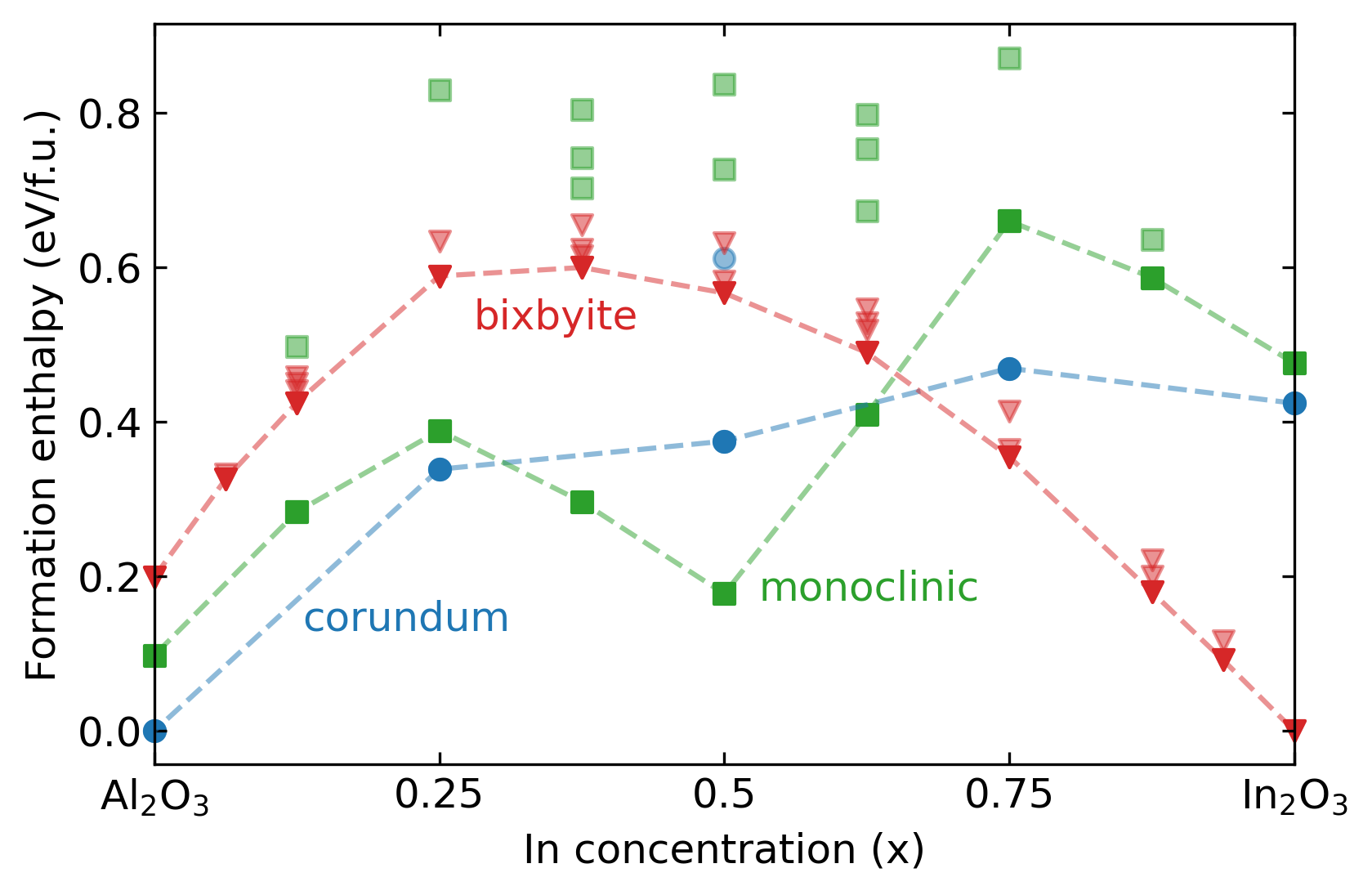}%
	\caption{\label{fig:formation}
		Enthalpy of formation per formula unit (f.u.) as function of In content for the most favorable configurations for each crystal structure. The colors and symbols represent each of the structures: blue circles represent corundum, red triangles represent bixbyite, and green squares represent monoclinic. The lowest energy configurations are given by solid symbols and the higher energy structures are represented by the semi-transparent symbols. The dashed lines connect each of the lowest energy points with straight lines and are shown as a visual aid. }
\end{figure}

Figure~\ref{fig:formation} shows the formation enthalpies as a function of In concentration for the three different crystal structures, with the corundum structures represented by blue circles, the bixbyite structures by red triangles, and the monoclinic structures by green squares. The lowest energy structures are shown with solid symbols and higher energy configurations are indicated by semi-transparent symbols. Consistent with the ground state structures of Al$_2$O$_3$ and In$_2$O$_3$, corundum structures have the lowest formation enthalpy for In concentrations of $x=0.25$ and smaller. For larger In concentrations ($x=0.75$ and greater), bixbyite structures are energetically favored. For concentrations around $x=0.5$ the monoclinic structure is the most energetically favored. This indicates that 50-50 InAlO$_3$ alloys will prefer monoclinic crystal structures, and not the crystal structures of their ground states (bixbyite or corundum). The reason behind this stabilization is that for these concentrations, ordered alloys, where all the In atoms occupy octahedral positions and all the Al atoms occupy tetrahedral positions, are formed. The bonding environments around Al and In closely resemble bonding environments in their bulk structures, lowering the energy of these structures. Note that a similar stabilization of ordered monoclinic structures has been observed in alloys of Ga$_2$O$_3$ with Al$_2$O$_3$~\cite{Peelaers2018,Wang2018b, Kim2021b} and of Ga$_2$O$_3$ with In$_2$O$_3$~\cite{Peelaers2015,Wouters2020}. Unordered alloys (see the semi-transparent symbols) are much higher in energy.

\begin{figure}[tb]
\includegraphics[width=0.95\columnwidth]{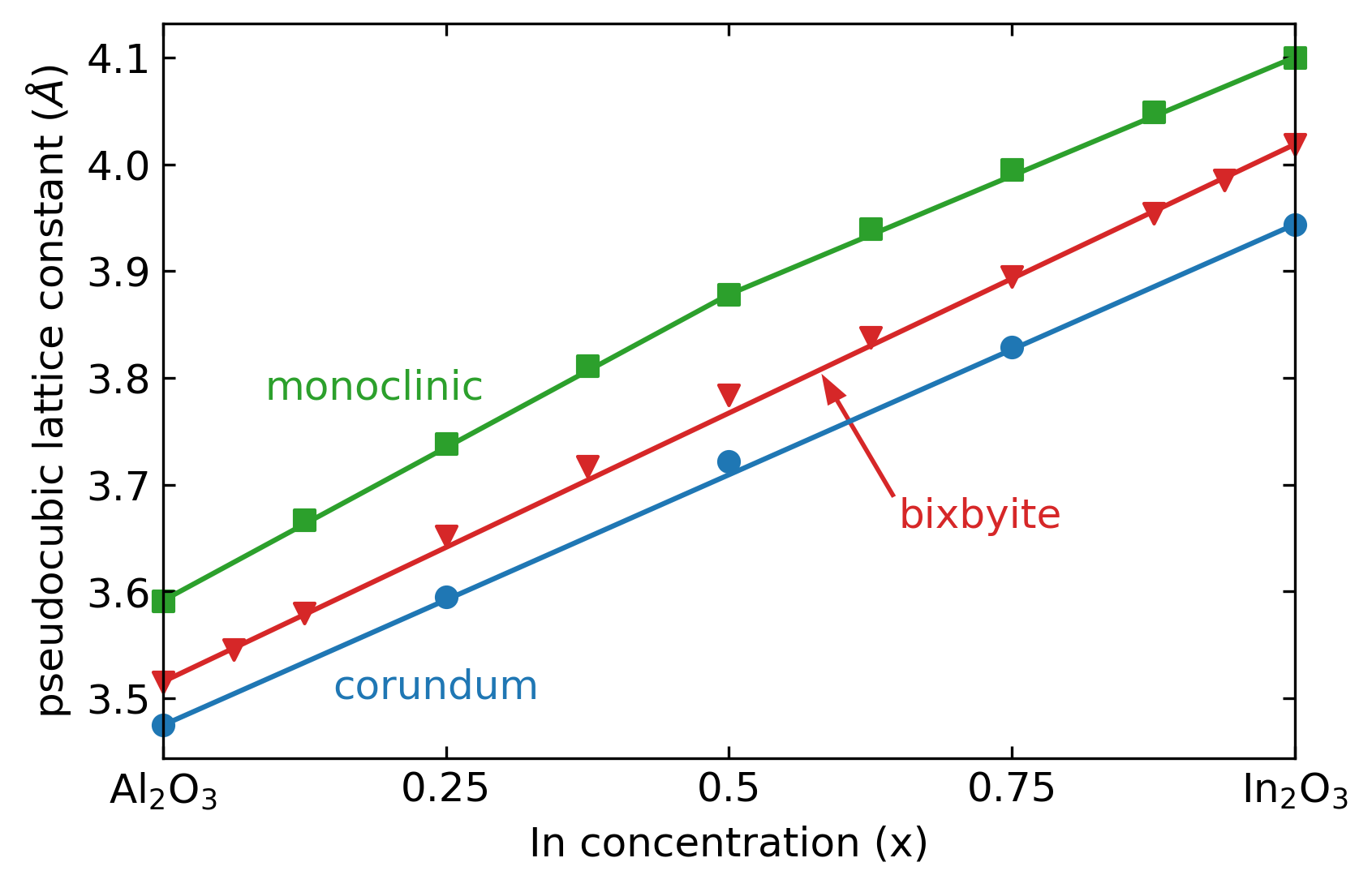}
\caption{\label{fig:lattice} Pseudocubic lattice constant as a function of In content for the lowest enthalpy corundum (blue), bixbyite (red), and monoclinic (green) structures given by Fig.~\ref{fig:formation}.  The solid lines are linear interpolations. }
\end{figure}

For each of the lowest enthalpy configurations, we plot in Fig.~\ref{fig:lattice} the pseudocubic lattice constant (the cube root of the volume) as a function of In content ($x$) for each of the crystal structures. We use the pseudocubic lattice constant to compare between the different crystal structures, as it is a single number that does not depend on the crystal symmetry (as compared to the three lattice constants and angles). Independent of the crystal structure, the pseudocubic lattice constant increases with the In concentration, consistent with the oxygen bond lengths in the ground state structures: The In-O bond in bixbyite In$_2$O$_3$ is longer than the Al-O bond in $\alpha$-Al$_2$O$_3$. The increase in the pseudocubic lattice constant is linear as In concentration increases, following Vegard's law. However, the monoclinic structure shows a kink at $x=0.5$, with a different linear slope before and after that concentration. The reason for this is that a different cation sublattice is being occupied with In: For In concentrations smaller than $x=0.5$, only octahedral sites are occupied with In atoms. Since only 50\% of the sites in the monoclinic structure are octahedral [see Fig.~\ref{fig:structures}(c)], the tetrahedral sites start to be filled after $x=0.5$, giving rise to a different slope.

Of particular interest is the 25\% In monoclinic structure. It has monoclinic lattice constants of $a$=3.03 \AA, $b$=12.22 \AA, $c$=5.79 \AA, and $\beta$=102.88$^\circ$.
These lattice constants are quite close to those of monoclinic Ga$_2$O$_3$ ($a$=3.04 \AA, $b$=12.25 \AA, $c$=5.79 \AA, and $\beta$=103.82$^\circ$). Creating heterostructures of monoclinic Ga$_2$O$_3$ and (In$_{0.25}$Al$_{0.75}$)$_2$O$_3$ would therefore introduce minimal strain. Note that the monoclinic structure has a slightly larger enthalpy of formation (0.05 eV/f.u.) compared to the corundum structure. This enthalpy difference is sufficiently small that growth kinetics can easily overcome it, especially when growing on a monoclinic Ga$_2$O$_3$ substrate.

\begin{figure}[tb]
\includegraphics[width=0.95\columnwidth]{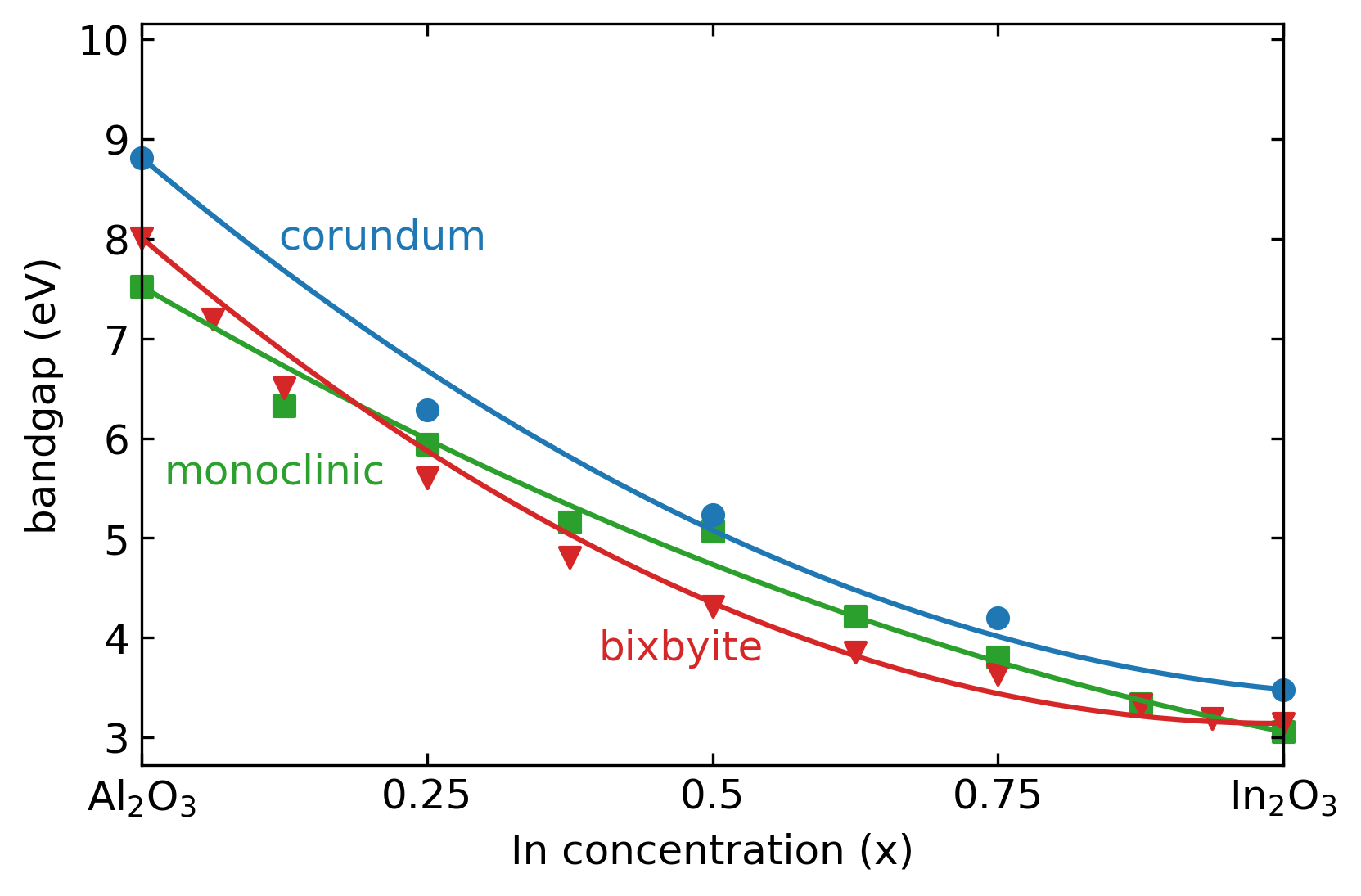}
\caption{\label{fig:gap} Direct bandgap at $\Gamma$ as a function of In content in each of the crystal structures considered. The corundum structures are represented by the blue circles, the bixbyite structures by the red triangles, and the monoclinic structures by the green squares. The solid lines are parabolic fits to Eq.(\ref{eq:bowing}) with $b=4.27$ eV or corundum, $b$=4.9 eV for bixbyite, and $b=2.22$ eV for monoclinic. }
\end{figure}

Next we investigate how the bandgap changes with increasing In concentration, as shown in Fig.~\ref{fig:gap}. For each of the crystal structures, the bandgap decreases as In content increases, which is commensurate with In$_2$O$_3$ possessing a smaller bandgap than Al$_2$O$_3$. This decrease is not linear, which we can characterize using the bowing parameter $b$, obtained by fitting our data to
\begin{align}
E_{g}(x) =  (1{\rm-}x)E_{g}[{\rm Al_2O_3}] +xE_{g}[{\rm In_2O_3}] - bx(1{\rm-}x) \, ,\label{eq:bowing}
\end{align}
where $E_{g}[{\rm Al_2O_3}]$ and $E_{g}[{\rm In_2O_3}]$ are the bandgaps of the 100\% Al structure and the 100\% In structure respectively.
From Eq. (\ref{eq:bowing}), the bowing parameter for each of the crystal structures is calculated as $b=2.22$ eV for monoclinic, $b=4.27$ eV for corundum, and $b=4.9$ eV for bixbyite. This indicates that the bandgaps of the monoclinic structures exhibit less bowing (so are more linear) compared to the bixbyite and corundum structures. The fitted curves are also shown in Fig.~\ref{fig:gap}.

While the values shown by Fig.~\ref{fig:gap} are for the direct bandgap at $\Gamma$, many of the structures have an indirect bandgap. For the monoclinic structures, the bandgap is indirect for all concentrations of In, with monoclinic In$_2$O$_3$ possessing an indirect gap of 3.04 eV (0.02 eV smaller than the direct gap) and monoclinic Al$_2$O$_3$ an indirect gap of 7.21 eV (0.31 eV difference). The bandgap is direct for the corundum In$_2$O$_3$ and Al$_2$O$_3$ endpoints, but becomes indirect for the intermediate concentrations. Fitting Eq. (\ref{eq:bowing}) to the indirect band gaps, gives slightly different values for the bowing, with $b=1.96$ eV for monoclinic, $b=4.91$ eV for corundum, and $b=5.6$ eV for bixbyite. 

For the 25\% In alloy of interest, the direct bandgap is 5.94 eV and the indirect bandgap is 5.70 eV. This is larger than the bandgap of Ga$_2$O$_3$, indicating that confinement of carriers in Ga$_2$O$_3$ might be possible.

\begin{figure}[tb]
	\includegraphics[width=0.7\columnwidth]{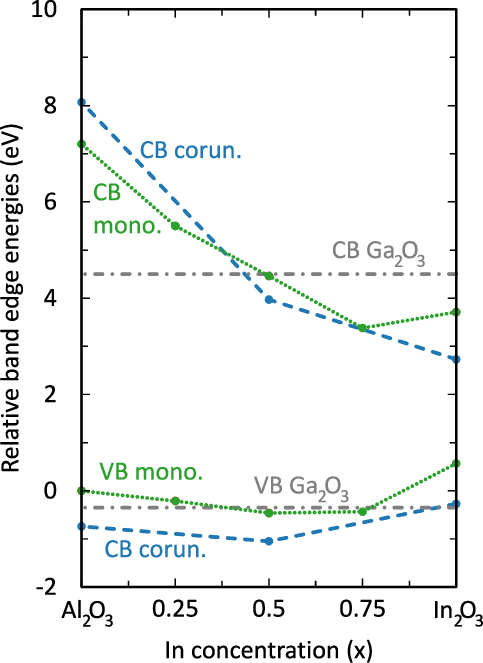} 
	\caption{\label{fig:alignment} The absolute energies of the valence- and conduction-band edges (in eV, relative to vacuum), as determined with slab calculations for alloys of corundum (blue dashed line) and monoclinic (green dotted line) (In$_{1-x}$Al$_{x}$)$_2$O$_3$. }
\end{figure}

To ascertain whether the difference in bandgaps leads to a beneficial conduction-band offset, we calculated the alignment of the valence- and conduction-band edges by relating them to the vacuum level (which forms a common reference energy). To do so, we constructed different slabs, with at least 25~\AA~of oxide and 25~\AA~of vacuum.
We then use the planar-averaged electrostatic potential in the direction perpendicular to the surface~\cite{Franciosi96} to obtain the alignment. The results are shown in Fig.~\ref{fig:alignment}, with all energies relative to the valence-band maximum (VBM) of monoclinic Al$_2$O$_3$, which is set to zero energy. 

We find that most of the bandgap change in the (In$_{0.25}$Al$_{0.75}$)$_2$O$_3$ alloy is reflected in the conduction band, which steadily decreases as a function of indium content. At 50\% alloy content, the band positions are quite similar to those of monoclinic Ga$_2$O$_3$. We also note that the band alignment between monoclinic and corundum Al$_2$O$_3$ is within 0.1 eV of prior studies~\cite{Peelaers2018}. Most importantly, we find that the (In$_{0.25}$Al$_{0.75}$)$_2$O$_3$ monoclinic alloy has a conduction-band offset of 1 eV, so that any charge carriers in a heterostructure with Ga$_2$O$_3$ should localize in the conduction band of Ga$_2$O$_3$. Combined with the lattice matching, this makes (In$_{0.25}$Al$_{0.75}$)$_2$O$_3$ the ideal alloy to use in device applications.

In summary, we used hybrid density functional theory to design a monoclinic alloy that lattice matches monoclinic Ga$_2$O$_3$, while providing a large conduction-band offset. To do so, we investigated the structural and electronic properties of~\inalo~alloys. We considered bixbyite, corundum, and monoclinic crystal structures. Lattice constants change linearly as a function of In concentration, except for monoclinic, which exhibits a change in slope at the 50\% In alloy. Bandgaps vary non-linearly with In concentration with bixbyite and corundum structures exhibiting more non-linearity (bowing parameters $b$ of 4.9 and 4.27 eV) compared to monoclinic alloys ($b$=2.22 eV). The 25\% In-content monoclinic alloy (In$_{0.25}$Al$_{0.75})_{2}$O$_3$ is a perfect lattice match to monoclinic Ga$_2$O$_3$ and also has a 1 eV conduction-band offset, making it the ideal alloy to use for heterostructures with Ga$_2$O$_3$.

\section{Acknowledgments*}

We thank Hari Nair for helpful discussions. JLL was supported by the Office of Naval Research (ONR) Basic Research Program. SS was supported by the University of Kansas General Research Fund (2151089). Computational resources were provided by Bridges-2 at Pittsburgh Supercomputing Center through allocation DMR200079 from the Advanced Cyberinfrastructure Coordination Ecosystem: Services \& Support (ACCESS) program, which is supported by National Science Foundation grants \#2138259, \#2138286, \#2138307, \#2137603, and \#2138296, by the DoD Major Shared Resource Center at AFRL, and by the University of Kansas Center for Research Computing (CRC), including the BigJay Cluster resource funded through NSF Grant MRI-2117449.

\section{Data Availability Statement*}
The data that support the findings of this study are available from the corresponding author upon reasonable request.
\bibliography{inalo}

\begin{thebibliography}{42}%
\makeatletter
\providecommand \@ifxundefined [1]{%
 \@ifx{#1\undefined}
}%
\providecommand \@ifnum [1]{%
 \ifnum #1\expandafter \@firstoftwo
 \else \expandafter \@secondoftwo
 \fi
}%
\providecommand \@ifx [1]{%
 \ifx #1\expandafter \@firstoftwo
 \else \expandafter \@secondoftwo
 \fi
}%
\providecommand \natexlab [1]{#1}%
\providecommand \enquote  [1]{``#1''}%
\providecommand \bibnamefont  [1]{#1}%
\providecommand \bibfnamefont [1]{#1}%
\providecommand \citenamefont [1]{#1}%
\providecommand \href@noop [0]{\@secondoftwo}%
\providecommand \href [0]{\begingroup \@sanitize@url \@href}%
\providecommand \@href[1]{\@@startlink{#1}\@@href}%
\providecommand \@@href[1]{\endgroup#1\@@endlink}%
\providecommand \@sanitize@url [0]{\catcode `\\12\catcode `\$12\catcode
  `\&12\catcode `\#12\catcode `\^12\catcode `\_12\catcode `\%12\relax}%
\providecommand \@@startlink[1]{}%
\providecommand \@@endlink[0]{}%
\providecommand \url  [0]{\begingroup\@sanitize@url \@url }%
\providecommand \@url [1]{\endgroup\@href {#1}{\urlprefix }}%
\providecommand \urlprefix  [0]{URL }%
\providecommand \Eprint [0]{\href }%
\providecommand \doibase [0]{http://dx.doi.org/}%
\providecommand \selectlanguage [0]{\@gobble}%
\providecommand \bibinfo  [0]{\@secondoftwo}%
\providecommand \bibfield  [0]{\@secondoftwo}%
\providecommand \translation [1]{[#1]}%
\providecommand \BibitemOpen [0]{}%
\providecommand \bibitemStop [0]{}%
\providecommand \bibitemNoStop [0]{.\EOS\space}%
\providecommand \EOS [0]{\spacefactor3000\relax}%
\providecommand \BibitemShut  [1]{\csname bibitem#1\endcsname}%
\let\auto@bib@innerbib\@empty
\bibitem [{\citenamefont {Tippins}(1965)}]{Tippins1965}%
  \BibitemOpen
  \bibfield  {author} {\bibinfo {author} {\bibfnamefont {H.~H.}\ \bibnamefont
  {Tippins}},\ }\href {\doibase 10.1103/PhysRev.140.A316} {\bibfield  {journal}
  {\bibinfo  {journal} {Phys. Rev.}\ }\textbf {\bibinfo {volume} {140}},\
  \bibinfo {pages} {A316} (\bibinfo {year} {1965})}\BibitemShut {NoStop}%
\bibitem [{\citenamefont {Matsumoto}\ \emph
  {et~al.}(1974{\natexlab{a}})\citenamefont {Matsumoto}, \citenamefont {Aoki},
  \citenamefont {Kinoshita},\ and\ \citenamefont {Aono}}]{Matsumoto1974}%
  \BibitemOpen
  \bibfield  {author} {\bibinfo {author} {\bibfnamefont {T.}~\bibnamefont
  {Matsumoto}}, \bibinfo {author} {\bibfnamefont {M.}~\bibnamefont {Aoki}},
  \bibinfo {author} {\bibfnamefont {A.}~\bibnamefont {Kinoshita}}, \ and\
  \bibinfo {author} {\bibfnamefont {T.}~\bibnamefont {Aono}},\ }\href {\doibase
  10.1143/JJAP.13.737} {\bibfield  {journal} {\bibinfo  {journal} {Jpn. J.
  Appl. Phys.}\ }\textbf {\bibinfo {volume} {13}},\ \bibinfo {pages} {737}
  (\bibinfo {year} {1974}{\natexlab{a}})}\BibitemShut {NoStop}%
\bibitem [{\citenamefont {Suzuki}\ \emph {et~al.}(2009)\citenamefont {Suzuki},
  \citenamefont {Nakagomi}, \citenamefont {Kokubun}, \citenamefont {Arai},\
  and\ \citenamefont {Ohira}}]{Suzuki2009}%
  \BibitemOpen
  \bibfield  {author} {\bibinfo {author} {\bibfnamefont {R.}~\bibnamefont
  {Suzuki}}, \bibinfo {author} {\bibfnamefont {S.}~\bibnamefont {Nakagomi}},
  \bibinfo {author} {\bibfnamefont {Y.}~\bibnamefont {Kokubun}}, \bibinfo
  {author} {\bibfnamefont {N.}~\bibnamefont {Arai}}, \ and\ \bibinfo {author}
  {\bibfnamefont {S.}~\bibnamefont {Ohira}},\ }\href {\doibase
  10.1063/1.3147197} {\bibfield  {journal} {\bibinfo  {journal} {Appl. Phys.
  Lett.}\ }\textbf {\bibinfo {volume} {94}},\ \bibinfo {pages} {222102}
  (\bibinfo {year} {2009})}\BibitemShut {NoStop}%
\bibitem [{\citenamefont {Hou}\ \emph {et~al.}(2021)\citenamefont {Hou},
  \citenamefont {Zou}, \citenamefont {Ding}, \citenamefont {Qin}, \citenamefont
  {Zhang}, \citenamefont {Ma}, \citenamefont {Tan}, \citenamefont {Yu},
  \citenamefont {Zhou}, \citenamefont {Zhao}, \citenamefont {Xu}, \citenamefont
  {Sun},\ and\ \citenamefont {Long}}]{Hou2021}%
  \BibitemOpen
  \bibfield  {author} {\bibinfo {author} {\bibfnamefont {X.}~\bibnamefont
  {Hou}}, \bibinfo {author} {\bibfnamefont {Y.}~\bibnamefont {Zou}}, \bibinfo
  {author} {\bibfnamefont {M.}~\bibnamefont {Ding}}, \bibinfo {author}
  {\bibfnamefont {Y.}~\bibnamefont {Qin}}, \bibinfo {author} {\bibfnamefont
  {Z.}~\bibnamefont {Zhang}}, \bibinfo {author} {\bibfnamefont
  {X.}~\bibnamefont {Ma}}, \bibinfo {author} {\bibfnamefont {P.}~\bibnamefont
  {Tan}}, \bibinfo {author} {\bibfnamefont {S.}~\bibnamefont {Yu}}, \bibinfo
  {author} {\bibfnamefont {X.}~\bibnamefont {Zhou}}, \bibinfo {author}
  {\bibfnamefont {X.}~\bibnamefont {Zhao}}, \bibinfo {author} {\bibfnamefont
  {G.}~\bibnamefont {Xu}}, \bibinfo {author} {\bibfnamefont {H.}~\bibnamefont
  {Sun}}, \ and\ \bibinfo {author} {\bibfnamefont {S.}~\bibnamefont {Long}},\
  }\href {\doibase 10.1088/1361-6463/abbb45} {\bibfield  {journal} {\bibinfo
  {journal} {J. Phys. D: Appl. Phys.}\ }\textbf {\bibinfo {volume} {54}},\
  \bibinfo {pages} {043001} (\bibinfo {year} {2021})}\BibitemShut {NoStop}%
\bibitem [{\citenamefont {Guo}\ \emph {et~al.}(2019)\citenamefont {Guo},
  \citenamefont {Guo}, \citenamefont {Chen}, \citenamefont {Wu}, \citenamefont
  {Li},\ and\ \citenamefont {Tang}}]{Guo2019}%
  \BibitemOpen
  \bibfield  {author} {\bibinfo {author} {\bibfnamefont {D.}~\bibnamefont
  {Guo}}, \bibinfo {author} {\bibfnamefont {Q.}~\bibnamefont {Guo}}, \bibinfo
  {author} {\bibfnamefont {Z.}~\bibnamefont {Chen}}, \bibinfo {author}
  {\bibfnamefont {Z.}~\bibnamefont {Wu}}, \bibinfo {author} {\bibfnamefont
  {P.}~\bibnamefont {Li}}, \ and\ \bibinfo {author} {\bibfnamefont
  {W.}~\bibnamefont {Tang}},\ }\href {\doibase 10.1016/j.mtphys.2019.100157}
  {\bibfield  {journal} {\bibinfo  {journal} {Mater. Today Phys.}\ }\textbf
  {\bibinfo {volume} {11}},\ \bibinfo {pages} {100157} (\bibinfo {year}
  {2019})}\BibitemShut {NoStop}%
\bibitem [{\citenamefont {Zou}\ \emph {et~al.}(2014)\citenamefont {Zou},
  \citenamefont {Zhang}, \citenamefont {Liu}, \citenamefont {Hu}, \citenamefont
  {Sang}, \citenamefont {Liao},\ and\ \citenamefont {Zhang}}]{Zou2014}%
  \BibitemOpen
  \bibfield  {author} {\bibinfo {author} {\bibfnamefont {R.}~\bibnamefont
  {Zou}}, \bibinfo {author} {\bibfnamefont {Z.}~\bibnamefont {Zhang}}, \bibinfo
  {author} {\bibfnamefont {Q.}~\bibnamefont {Liu}}, \bibinfo {author}
  {\bibfnamefont {J.}~\bibnamefont {Hu}}, \bibinfo {author} {\bibfnamefont
  {L.}~\bibnamefont {Sang}}, \bibinfo {author} {\bibfnamefont {M.}~\bibnamefont
  {Liao}}, \ and\ \bibinfo {author} {\bibfnamefont {W.}~\bibnamefont {Zhang}},\
  }\href {\doibase 10.1002/smll.201302705} {\bibfield  {journal} {\bibinfo
  {journal} {Small}\ }\textbf {\bibinfo {volume} {10}},\ \bibinfo {pages}
  {1848} (\bibinfo {year} {2014})}\BibitemShut {NoStop}%
\bibitem [{\citenamefont {Zhang}\ \emph {et~al.}(2020)\citenamefont {Zhang},
  \citenamefont {Shi}, \citenamefont {Qi}, \citenamefont {Chen},\ and\
  \citenamefont {Zhang}}]{Zhang2020}%
  \BibitemOpen
  \bibfield  {author} {\bibinfo {author} {\bibfnamefont {J.}~\bibnamefont
  {Zhang}}, \bibinfo {author} {\bibfnamefont {J.}~\bibnamefont {Shi}}, \bibinfo
  {author} {\bibfnamefont {D.-C.}\ \bibnamefont {Qi}}, \bibinfo {author}
  {\bibfnamefont {L.}~\bibnamefont {Chen}}, \ and\ \bibinfo {author}
  {\bibfnamefont {K.~H.~L.}\ \bibnamefont {Zhang}},\ }\href {\doibase
  10.1063/1.5142999} {\bibfield  {journal} {\bibinfo  {journal} {APL Mater.}\
  }\textbf {\bibinfo {volume} {8}},\ \bibinfo {pages} {020906} (\bibinfo {year}
  {2020})}\BibitemShut {NoStop}%
\bibitem [{\citenamefont {Pearton}\ \emph
  {et~al.}(2018{\natexlab{a}})\citenamefont {Pearton}, \citenamefont {Ren},
  \citenamefont {Tadjer},\ and\ \citenamefont {Kim}}]{Pearton2018b}%
  \BibitemOpen
  \bibfield  {author} {\bibinfo {author} {\bibfnamefont {S.~J.}\ \bibnamefont
  {Pearton}}, \bibinfo {author} {\bibfnamefont {F.}~\bibnamefont {Ren}},
  \bibinfo {author} {\bibfnamefont {M.}~\bibnamefont {Tadjer}}, \ and\ \bibinfo
  {author} {\bibfnamefont {J.}~\bibnamefont {Kim}},\ }\href {\doibase
  10.1063/1.5062841} {\bibfield  {journal} {\bibinfo  {journal} {J. Appl.
  Phys.}\ }\textbf {\bibinfo {volume} {124}},\ \bibinfo {pages} {220901}
  (\bibinfo {year} {2018}{\natexlab{a}})}\BibitemShut {NoStop}%
\bibitem [{\citenamefont {Higashiwaki}\ \emph {et~al.}(2017)\citenamefont
  {Higashiwaki}, \citenamefont {Kuramata}, \citenamefont {Murakami},\ and\
  \citenamefont {Kumagai}}]{Higashiwaki2017}%
  \BibitemOpen
  \bibfield  {author} {\bibinfo {author} {\bibfnamefont {M.}~\bibnamefont
  {Higashiwaki}}, \bibinfo {author} {\bibfnamefont {A.}~\bibnamefont
  {Kuramata}}, \bibinfo {author} {\bibfnamefont {H.}~\bibnamefont {Murakami}},
  \ and\ \bibinfo {author} {\bibfnamefont {Y.}~\bibnamefont {Kumagai}},\ }\href
  {\doibase 10.1088/1361-6463/aa7aff} {\bibfield  {journal} {\bibinfo
  {journal} {J. Phys. D: Appl. Phys.}\ }\textbf {\bibinfo {volume} {50}},\
  \bibinfo {pages} {333002} (\bibinfo {year} {2017})}\BibitemShut {NoStop}%
\bibitem [{\citenamefont {Pearton}\ \emph
  {et~al.}(2018{\natexlab{b}})\citenamefont {Pearton}, \citenamefont {Yang},
  \citenamefont {Cary}, \citenamefont {Ren}, \citenamefont {Kim}, \citenamefont
  {Tadjer},\ and\ \citenamefont {Mastro}}]{Pearton2018a}%
  \BibitemOpen
  \bibfield  {author} {\bibinfo {author} {\bibfnamefont {S.~J.}\ \bibnamefont
  {Pearton}}, \bibinfo {author} {\bibfnamefont {J.}~\bibnamefont {Yang}},
  \bibinfo {author} {\bibfnamefont {P.~H.}\ \bibnamefont {Cary}}, \bibinfo
  {author} {\bibfnamefont {F.}~\bibnamefont {Ren}}, \bibinfo {author}
  {\bibfnamefont {J.}~\bibnamefont {Kim}}, \bibinfo {author} {\bibfnamefont
  {M.~J.}\ \bibnamefont {Tadjer}}, \ and\ \bibinfo {author} {\bibfnamefont
  {M.~A.}\ \bibnamefont {Mastro}},\ }\href {\doibase 10.1063/1.5006941}
  {\bibfield  {journal} {\bibinfo  {journal} {Appl. Phys. Rev.}\ }\textbf
  {\bibinfo {volume} {5}},\ \bibinfo {pages} {011301} (\bibinfo {year}
  {2018}{\natexlab{b}})}\BibitemShut {NoStop}%
\bibitem [{\citenamefont {Ahmadi}\ \emph {et~al.}(2017)\citenamefont {Ahmadi},
  \citenamefont {Koksaldi}, \citenamefont {Zheng}, \citenamefont {Mates},
  \citenamefont {Oshima}, \citenamefont {Mishra},\ and\ \citenamefont
  {Speck}}]{Ahmadi2017a}%
  \BibitemOpen
  \bibfield  {author} {\bibinfo {author} {\bibfnamefont {E.}~\bibnamefont
  {Ahmadi}}, \bibinfo {author} {\bibfnamefont {O.~S.}\ \bibnamefont
  {Koksaldi}}, \bibinfo {author} {\bibfnamefont {X.}~\bibnamefont {Zheng}},
  \bibinfo {author} {\bibfnamefont {T.}~\bibnamefont {Mates}}, \bibinfo
  {author} {\bibfnamefont {Y.}~\bibnamefont {Oshima}}, \bibinfo {author}
  {\bibfnamefont {U.~K.}\ \bibnamefont {Mishra}}, \ and\ \bibinfo {author}
  {\bibfnamefont {J.~S.}\ \bibnamefont {Speck}},\ }\href {\doibase
  10.7567/APEX.10.071101} {\bibfield  {journal} {\bibinfo  {journal} {Appl.
  Phys. Express}\ }\textbf {\bibinfo {volume} {10}},\ \bibinfo {pages} {071101}
  (\bibinfo {year} {2017})}\BibitemShut {NoStop}%
\bibitem [{\citenamefont {Krishnamoorthy}\ \emph {et~al.}(2017)\citenamefont
  {Krishnamoorthy}, \citenamefont {Xia}, \citenamefont {Joishi}, \citenamefont
  {Zhang}, \citenamefont {McGlone}, \citenamefont {Johnson}, \citenamefont
  {Brenner}, \citenamefont {Arehart}, \citenamefont {Hwang}, \citenamefont
  {Lodha},\ and\ \citenamefont {Rajan}}]{Krishnamoorthy2017}%
  \BibitemOpen
  \bibfield  {author} {\bibinfo {author} {\bibfnamefont {S.}~\bibnamefont
  {Krishnamoorthy}}, \bibinfo {author} {\bibfnamefont {Z.}~\bibnamefont {Xia}},
  \bibinfo {author} {\bibfnamefont {C.}~\bibnamefont {Joishi}}, \bibinfo
  {author} {\bibfnamefont {Y.}~\bibnamefont {Zhang}}, \bibinfo {author}
  {\bibfnamefont {J.}~\bibnamefont {McGlone}}, \bibinfo {author} {\bibfnamefont
  {J.}~\bibnamefont {Johnson}}, \bibinfo {author} {\bibfnamefont
  {M.}~\bibnamefont {Brenner}}, \bibinfo {author} {\bibfnamefont {A.~R.}\
  \bibnamefont {Arehart}}, \bibinfo {author} {\bibfnamefont {J.}~\bibnamefont
  {Hwang}}, \bibinfo {author} {\bibfnamefont {S.}~\bibnamefont {Lodha}}, \ and\
  \bibinfo {author} {\bibfnamefont {S.}~\bibnamefont {Rajan}},\ }\href
  {\doibase 10.1063/1.4993569} {\bibfield  {journal} {\bibinfo  {journal}
  {Appl. Phys. Lett.}\ }\textbf {\bibinfo {volume} {111}},\ \bibinfo {pages}
  {023502} (\bibinfo {year} {2017})}\BibitemShut {NoStop}%
\bibitem [{\citenamefont {Zhang}\ \emph {et~al.}(2018)\citenamefont {Zhang},
  \citenamefont {Neal}, \citenamefont {Xia}, \citenamefont {Joishi},
  \citenamefont {Johnson}, \citenamefont {Zheng}, \citenamefont {Bajaj},
  \citenamefont {Brenner}, \citenamefont {Dorsey}, \citenamefont {Chabak},
  \citenamefont {Jessen}, \citenamefont {Hwang}, \citenamefont {Mou},
  \citenamefont {Heremans},\ and\ \citenamefont {Rajan}}]{Zhang2018d}%
  \BibitemOpen
  \bibfield  {author} {\bibinfo {author} {\bibfnamefont {Y.}~\bibnamefont
  {Zhang}}, \bibinfo {author} {\bibfnamefont {A.}~\bibnamefont {Neal}},
  \bibinfo {author} {\bibfnamefont {Z.}~\bibnamefont {Xia}}, \bibinfo {author}
  {\bibfnamefont {C.}~\bibnamefont {Joishi}}, \bibinfo {author} {\bibfnamefont
  {J.~M.}\ \bibnamefont {Johnson}}, \bibinfo {author} {\bibfnamefont
  {Y.}~\bibnamefont {Zheng}}, \bibinfo {author} {\bibfnamefont
  {S.}~\bibnamefont {Bajaj}}, \bibinfo {author} {\bibfnamefont
  {M.}~\bibnamefont {Brenner}}, \bibinfo {author} {\bibfnamefont
  {D.}~\bibnamefont {Dorsey}}, \bibinfo {author} {\bibfnamefont
  {K.}~\bibnamefont {Chabak}}, \bibinfo {author} {\bibfnamefont
  {G.}~\bibnamefont {Jessen}}, \bibinfo {author} {\bibfnamefont
  {J.}~\bibnamefont {Hwang}}, \bibinfo {author} {\bibfnamefont
  {S.}~\bibnamefont {Mou}}, \bibinfo {author} {\bibfnamefont {J.~P.}\
  \bibnamefont {Heremans}}, \ and\ \bibinfo {author} {\bibfnamefont
  {S.}~\bibnamefont {Rajan}},\ }\href {\doibase 10.1063/1.5025704} {\bibfield
  {journal} {\bibinfo  {journal} {Appl. Phys. Lett.}\ }\textbf {\bibinfo
  {volume} {112}},\ \bibinfo {pages} {173502} (\bibinfo {year}
  {2018})}\BibitemShut {NoStop}%
\bibitem [{\citenamefont {Joishi}\ \emph {et~al.}(2019)\citenamefont {Joishi},
  \citenamefont {Zhang}, \citenamefont {Xia}, \citenamefont {Sun},
  \citenamefont {Arehart}, \citenamefont {Ringel}, \citenamefont {Lodha},\ and\
  \citenamefont {Rajan}}]{Joishi2019}%
  \BibitemOpen
  \bibfield  {author} {\bibinfo {author} {\bibfnamefont {C.}~\bibnamefont
  {Joishi}}, \bibinfo {author} {\bibfnamefont {Y.}~\bibnamefont {Zhang}},
  \bibinfo {author} {\bibfnamefont {Z.}~\bibnamefont {Xia}}, \bibinfo {author}
  {\bibfnamefont {W.}~\bibnamefont {Sun}}, \bibinfo {author} {\bibfnamefont
  {A.~R.}\ \bibnamefont {Arehart}}, \bibinfo {author} {\bibfnamefont
  {S.}~\bibnamefont {Ringel}}, \bibinfo {author} {\bibfnamefont
  {S.}~\bibnamefont {Lodha}}, \ and\ \bibinfo {author} {\bibfnamefont
  {S.}~\bibnamefont {Rajan}},\ }\href {\doibase 10.1109/LED.2019.2921116}
  {\bibfield  {journal} {\bibinfo  {journal} {IEEE Electron Device Lett.}\
  }\textbf {\bibinfo {volume} {40}},\ \bibinfo {pages} {1241} (\bibinfo {year}
  {2019})}\BibitemShut {NoStop}%
\bibitem [{\citenamefont {Okumura}\ \emph {et~al.}(2019)\citenamefont
  {Okumura}, \citenamefont {Kato}, \citenamefont {Oshima},\ and\ \citenamefont
  {Palacios}}]{Okumura2019}%
  \BibitemOpen
  \bibfield  {author} {\bibinfo {author} {\bibfnamefont {H.}~\bibnamefont
  {Okumura}}, \bibinfo {author} {\bibfnamefont {Y.}~\bibnamefont {Kato}},
  \bibinfo {author} {\bibfnamefont {T.}~\bibnamefont {Oshima}}, \ and\ \bibinfo
  {author} {\bibfnamefont {T.}~\bibnamefont {Palacios}},\ }\href {\doibase
  10.7567/1347-4065/ab002b} {\bibfield  {journal} {\bibinfo  {journal} {Jpn. J.
  Appl. Phys.}\ }\textbf {\bibinfo {volume} {58}},\ \bibinfo {pages} {SBBD12}
  (\bibinfo {year} {2019})}\BibitemShut {NoStop}%
\bibitem [{\citenamefont {Chatterjee}\ \emph {et~al.}(2020)\citenamefont
  {Chatterjee}, \citenamefont {Song}, \citenamefont {Lundh}, \citenamefont
  {Zhang}, \citenamefont {Xia}, \citenamefont {Islam}, \citenamefont {Leach},
  \citenamefont {McGray}, \citenamefont {Ranga}, \citenamefont
  {Krishnamoorthy}, \citenamefont {Haque}, \citenamefont {Rajan},\ and\
  \citenamefont {Choi}}]{Chatterjee2020}%
  \BibitemOpen
  \bibfield  {author} {\bibinfo {author} {\bibfnamefont {B.}~\bibnamefont
  {Chatterjee}}, \bibinfo {author} {\bibfnamefont {Y.}~\bibnamefont {Song}},
  \bibinfo {author} {\bibfnamefont {J.~S.}\ \bibnamefont {Lundh}}, \bibinfo
  {author} {\bibfnamefont {Y.}~\bibnamefont {Zhang}}, \bibinfo {author}
  {\bibfnamefont {Z.}~\bibnamefont {Xia}}, \bibinfo {author} {\bibfnamefont
  {Z.}~\bibnamefont {Islam}}, \bibinfo {author} {\bibfnamefont
  {J.}~\bibnamefont {Leach}}, \bibinfo {author} {\bibfnamefont
  {C.}~\bibnamefont {McGray}}, \bibinfo {author} {\bibfnamefont
  {P.}~\bibnamefont {Ranga}}, \bibinfo {author} {\bibfnamefont
  {S.}~\bibnamefont {Krishnamoorthy}}, \bibinfo {author} {\bibfnamefont
  {A.}~\bibnamefont {Haque}}, \bibinfo {author} {\bibfnamefont
  {S.}~\bibnamefont {Rajan}}, \ and\ \bibinfo {author} {\bibfnamefont
  {S.}~\bibnamefont {Choi}},\ }\href {\doibase 10.1063/5.0021275} {\bibfield
  {journal} {\bibinfo  {journal} {Appl. Phys. Lett.}\ }\textbf {\bibinfo
  {volume} {117}},\ \bibinfo {pages} {153501} (\bibinfo {year}
  {2020})}\BibitemShut {NoStop}%
\bibitem [{\citenamefont {Vaidya}\ \emph {et~al.}(2021)\citenamefont {Vaidya},
  \citenamefont {Saha},\ and\ \citenamefont {Singisetti}}]{Vaidya2021}%
  \BibitemOpen
  \bibfield  {author} {\bibinfo {author} {\bibfnamefont {A.}~\bibnamefont
  {Vaidya}}, \bibinfo {author} {\bibfnamefont {C.~N.}\ \bibnamefont {Saha}}, \
  and\ \bibinfo {author} {\bibfnamefont {U.}~\bibnamefont {Singisetti}},\
  }\href {\doibase 10.1109/LED.2021.3104256} {\bibfield  {journal} {\bibinfo
  {journal} {IEEE Electron Device Lett.}\ }\textbf {\bibinfo {volume} {42}},\
  \bibinfo {pages} {1444} (\bibinfo {year} {2021})}\BibitemShut {NoStop}%
\bibitem [{\citenamefont {Tadjer}\ \emph {et~al.}(2021)\citenamefont {Tadjer},
  \citenamefont {Sasaki}, \citenamefont {Wakimoto}, \citenamefont {Anderson},
  \citenamefont {Mastro}, \citenamefont {Gallagher}, \citenamefont {Jacobs},
  \citenamefont {Mock}, \citenamefont {Koehler}, \citenamefont {Ebrish},
  \citenamefont {Hobart},\ and\ \citenamefont {Kuramata}}]{Tadjer2021}%
  \BibitemOpen
  \bibfield  {author} {\bibinfo {author} {\bibfnamefont {M.~J.}\ \bibnamefont
  {Tadjer}}, \bibinfo {author} {\bibfnamefont {K.}~\bibnamefont {Sasaki}},
  \bibinfo {author} {\bibfnamefont {D.}~\bibnamefont {Wakimoto}}, \bibinfo
  {author} {\bibfnamefont {T.~J.}\ \bibnamefont {Anderson}}, \bibinfo {author}
  {\bibfnamefont {M.~A.}\ \bibnamefont {Mastro}}, \bibinfo {author}
  {\bibfnamefont {J.~C.}\ \bibnamefont {Gallagher}}, \bibinfo {author}
  {\bibfnamefont {A.~G.}\ \bibnamefont {Jacobs}}, \bibinfo {author}
  {\bibfnamefont {A.~L.}\ \bibnamefont {Mock}}, \bibinfo {author}
  {\bibfnamefont {A.~D.}\ \bibnamefont {Koehler}}, \bibinfo {author}
  {\bibfnamefont {M.}~\bibnamefont {Ebrish}}, \bibinfo {author} {\bibfnamefont
  {K.~D.}\ \bibnamefont {Hobart}}, \ and\ \bibinfo {author} {\bibfnamefont
  {A.}~\bibnamefont {Kuramata}},\ }\href {\doibase 10.1116/6.0000932}
  {\bibfield  {journal} {\bibinfo  {journal} {J. Vac. Sci. Technol. A}\
  }\textbf {\bibinfo {volume} {39}},\ \bibinfo {pages} {033402} (\bibinfo
  {year} {2021})}\BibitemShut {NoStop}%
\bibitem [{\citenamefont {Peelaers}\ \emph {et~al.}(2018)\citenamefont
  {Peelaers}, \citenamefont {Varley}, \citenamefont {Speck},\ and\
  \citenamefont {Van~de Walle}}]{Peelaers2018}%
  \BibitemOpen
  \bibfield  {author} {\bibinfo {author} {\bibfnamefont {H.}~\bibnamefont
  {Peelaers}}, \bibinfo {author} {\bibfnamefont {J.~B.}\ \bibnamefont
  {Varley}}, \bibinfo {author} {\bibfnamefont {J.~S.}\ \bibnamefont {Speck}}, \
  and\ \bibinfo {author} {\bibfnamefont {C.~G.}\ \bibnamefont {Van~de Walle}},\
  }\href {\doibase 10.1063/1.5036991} {\bibfield  {journal} {\bibinfo
  {journal} {Appl. Phys. Lett.}\ }\textbf {\bibinfo {volume} {112}},\ \bibinfo
  {pages} {242101} (\bibinfo {year} {2018})}\BibitemShut {NoStop}%
\bibitem [{\citenamefont {Peelaers}\ \emph {et~al.}(2015)\citenamefont
  {Peelaers}, \citenamefont {Steiauf}, \citenamefont {Varley}, \citenamefont
  {Janotti},\ and\ \citenamefont {{Van de Walle}}}]{Peelaers2015}%
  \BibitemOpen
  \bibfield  {author} {\bibinfo {author} {\bibfnamefont {H.}~\bibnamefont
  {Peelaers}}, \bibinfo {author} {\bibfnamefont {D.}~\bibnamefont {Steiauf}},
  \bibinfo {author} {\bibfnamefont {J.~B.}\ \bibnamefont {Varley}}, \bibinfo
  {author} {\bibfnamefont {A.}~\bibnamefont {Janotti}}, \ and\ \bibinfo
  {author} {\bibfnamefont {C.~G.}\ \bibnamefont {{Van de Walle}}},\ }\href
  {\doibase 10.1103/PhysRevB.92.085206} {\bibfield  {journal} {\bibinfo
  {journal} {Phys. Rev. B}\ }\textbf {\bibinfo {volume} {92}},\ \bibinfo
  {pages} {085206} (\bibinfo {year} {2015})}\BibitemShut {NoStop}%
\bibitem [{\citenamefont {Seacat}\ \emph {et~al.}(2020)\citenamefont {Seacat},
  \citenamefont {Lyons},\ and\ \citenamefont {Peelaers}}]{Seacat2020}%
  \BibitemOpen
  \bibfield  {author} {\bibinfo {author} {\bibfnamefont {S.}~\bibnamefont
  {Seacat}}, \bibinfo {author} {\bibfnamefont {J.~L.}\ \bibnamefont {Lyons}}, \
  and\ \bibinfo {author} {\bibfnamefont {H.}~\bibnamefont {Peelaers}},\ }\href
  {\doibase 10.1063/5.0010354} {\bibfield  {journal} {\bibinfo  {journal}
  {Appl. Phys. Lett.}\ }\textbf {\bibinfo {volume} {116}},\ \bibinfo {pages}
  {232102} (\bibinfo {year} {2020})}\BibitemShut {NoStop}%
\bibitem [{\citenamefont {Seacat}\ \emph {et~al.}(2021)\citenamefont {Seacat},
  \citenamefont {Lyons},\ and\ \citenamefont {Peelaers}}]{Seacat2021}%
  \BibitemOpen
  \bibfield  {author} {\bibinfo {author} {\bibfnamefont {S.}~\bibnamefont
  {Seacat}}, \bibinfo {author} {\bibfnamefont {J.~L.}\ \bibnamefont {Lyons}}, \
  and\ \bibinfo {author} {\bibfnamefont {H.}~\bibnamefont {Peelaers}},\ }\href
  {\doibase 10.1063/5.0060801} {\bibfield  {journal} {\bibinfo  {journal}
  {Appl. Phys. Lett.}\ }\textbf {\bibinfo {volume} {119}},\ \bibinfo {pages}
  {042104} (\bibinfo {year} {2021})}\BibitemShut {NoStop}%
\bibitem [{\citenamefont {Wang}\ \emph {et~al.}(2018)\citenamefont {Wang},
  \citenamefont {Li}, \citenamefont {Ni},\ and\ \citenamefont
  {Janotti}}]{Wang2018b}%
  \BibitemOpen
  \bibfield  {author} {\bibinfo {author} {\bibfnamefont {T.}~\bibnamefont
  {Wang}}, \bibinfo {author} {\bibfnamefont {W.}~\bibnamefont {Li}}, \bibinfo
  {author} {\bibfnamefont {C.}~\bibnamefont {Ni}}, \ and\ \bibinfo {author}
  {\bibfnamefont {A.}~\bibnamefont {Janotti}},\ }\href {\doibase
  10.1103/PhysRevApplied.10.011003} {\bibfield  {journal} {\bibinfo  {journal}
  {Phys. Rev. Applied}\ }\textbf {\bibinfo {volume} {10}},\ \bibinfo {pages}
  {011003} (\bibinfo {year} {2018})}\BibitemShut {NoStop}%
\bibitem [{\citenamefont {Kim}\ \emph {et~al.}(2021)\citenamefont {Kim},
  \citenamefont {Ko}, \citenamefont {Chung},\ and\ \citenamefont
  {Cho}}]{Kim2021b}%
  \BibitemOpen
  \bibfield  {author} {\bibinfo {author} {\bibfnamefont {H.~W.}\ \bibnamefont
  {Kim}}, \bibinfo {author} {\bibfnamefont {H.}~\bibnamefont {Ko}}, \bibinfo
  {author} {\bibfnamefont {Y.-C.}\ \bibnamefont {Chung}}, \ and\ \bibinfo
  {author} {\bibfnamefont {S.~B.}\ \bibnamefont {Cho}},\ }\href {\doibase
  10.1016/j.jeurceramsoc.2020.08.067} {\bibfield  {journal} {\bibinfo
  {journal} {J. Eur. Ceram. Soc.}\ }\textbf {\bibinfo {volume} {41}},\ \bibinfo
  {pages} {611} (\bibinfo {year} {2021})}\BibitemShut {NoStop}%
\bibitem [{\citenamefont {Wouters}\ \emph {et~al.}(2020)\citenamefont
  {Wouters}, \citenamefont {Sutton}, \citenamefont {Ghiringhelli},
  \citenamefont {Markurt}, \citenamefont {Schewski}, \citenamefont {Hassa},
  \citenamefont {{von Wenckstern}}, \citenamefont {Grundmann}, \citenamefont
  {Scheffler},\ and\ \citenamefont {Albrecht}}]{Wouters2020}%
  \BibitemOpen
  \bibfield  {author} {\bibinfo {author} {\bibfnamefont {C.}~\bibnamefont
  {Wouters}}, \bibinfo {author} {\bibfnamefont {C.}~\bibnamefont {Sutton}},
  \bibinfo {author} {\bibfnamefont {L.~M.}\ \bibnamefont {Ghiringhelli}},
  \bibinfo {author} {\bibfnamefont {T.}~\bibnamefont {Markurt}}, \bibinfo
  {author} {\bibfnamefont {R.}~\bibnamefont {Schewski}}, \bibinfo {author}
  {\bibfnamefont {A.}~\bibnamefont {Hassa}}, \bibinfo {author} {\bibfnamefont
  {H.}~\bibnamefont {{von Wenckstern}}}, \bibinfo {author} {\bibfnamefont
  {M.}~\bibnamefont {Grundmann}}, \bibinfo {author} {\bibfnamefont
  {M.}~\bibnamefont {Scheffler}}, \ and\ \bibinfo {author} {\bibfnamefont
  {M.}~\bibnamefont {Albrecht}},\ }\href {\doibase
  10.1103/PhysRevMaterials.4.125001} {\bibfield  {journal} {\bibinfo  {journal}
  {Phys. Rev. Mater.}\ }\textbf {\bibinfo {volume} {4}},\ \bibinfo {pages}
  {125001} (\bibinfo {year} {2020})}\BibitemShut {NoStop}%
\bibitem [{\citenamefont {Mu}\ \emph {et~al.}(2020)\citenamefont {Mu},
  \citenamefont {Wang}, \citenamefont {Peelaers},\ and\ \citenamefont {{Van de
  Walle}}}]{Mu2020}%
  \BibitemOpen
  \bibfield  {author} {\bibinfo {author} {\bibfnamefont {S.}~\bibnamefont
  {Mu}}, \bibinfo {author} {\bibfnamefont {M.}~\bibnamefont {Wang}}, \bibinfo
  {author} {\bibfnamefont {H.}~\bibnamefont {Peelaers}}, \ and\ \bibinfo
  {author} {\bibfnamefont {C.~G.}\ \bibnamefont {{Van de Walle}}},\ }\href
  {\doibase 10.1063/5.0019915} {\bibfield  {journal} {\bibinfo  {journal} {APL
  Mater.}\ }\textbf {\bibinfo {volume} {8}},\ \bibinfo {pages} {091105}
  (\bibinfo {year} {2020})}\BibitemShut {NoStop}%
\bibitem [{\citenamefont {Lundh}\ \emph {et~al.}(2023)\citenamefont {Lundh},
  \citenamefont {Huynh}, \citenamefont {Liao}, \citenamefont {Olsen},
  \citenamefont {Pan}, \citenamefont {Sasaki}, \citenamefont {Konishi},
  \citenamefont {Masten}, \citenamefont {Hite}, \citenamefont {Mastro},
  \citenamefont {Mahadik}, \citenamefont {Goorsky}, \citenamefont {Kuramata},
  \citenamefont {Hobart}, \citenamefont {Anderson},\ and\ \citenamefont
  {Tadjer}}]{Lundh2023}%
  \BibitemOpen
  \bibfield  {author} {\bibinfo {author} {\bibfnamefont {J.~S.}\ \bibnamefont
  {Lundh}}, \bibinfo {author} {\bibfnamefont {K.}~\bibnamefont {Huynh}},
  \bibinfo {author} {\bibfnamefont {M.}~\bibnamefont {Liao}}, \bibinfo {author}
  {\bibfnamefont {W.}~\bibnamefont {Olsen}}, \bibinfo {author} {\bibfnamefont
  {K.}~\bibnamefont {Pan}}, \bibinfo {author} {\bibfnamefont {K.}~\bibnamefont
  {Sasaki}}, \bibinfo {author} {\bibfnamefont {K.}~\bibnamefont {Konishi}},
  \bibinfo {author} {\bibfnamefont {H.~N.}\ \bibnamefont {Masten}}, \bibinfo
  {author} {\bibfnamefont {J.~K.}\ \bibnamefont {Hite}}, \bibinfo {author}
  {\bibfnamefont {M.~A.}\ \bibnamefont {Mastro}}, \bibinfo {author}
  {\bibfnamefont {N.~A.}\ \bibnamefont {Mahadik}}, \bibinfo {author}
  {\bibfnamefont {M.}~\bibnamefont {Goorsky}}, \bibinfo {author} {\bibfnamefont
  {A.}~\bibnamefont {Kuramata}}, \bibinfo {author} {\bibfnamefont {K.~D.}\
  \bibnamefont {Hobart}}, \bibinfo {author} {\bibfnamefont {T.~J.}\
  \bibnamefont {Anderson}}, \ and\ \bibinfo {author} {\bibfnamefont {M.~J.}\
  \bibnamefont {Tadjer}},\ }\href {\doibase 10.1063/5.0174682} {\bibfield
  {journal} {\bibinfo  {journal} {Appl. Phys. Lett.}\ }\textbf {\bibinfo
  {volume} {123}},\ \bibinfo {pages} {222104} (\bibinfo {year}
  {2023})}\BibitemShut {NoStop}%
\bibitem [{\citenamefont {Ohtsuki}\ and\ \citenamefont
  {Higashiwaki}(2023)}]{Ohtsuki2023}%
  \BibitemOpen
  \bibfield  {author} {\bibinfo {author} {\bibfnamefont {T.}~\bibnamefont
  {Ohtsuki}}\ and\ \bibinfo {author} {\bibfnamefont {M.}~\bibnamefont
  {Higashiwaki}},\ }\href {\doibase 10.1116/6.0002625} {\bibfield  {journal}
  {\bibinfo  {journal} {J. Vac. Sci. Technol. A}\ }\textbf {\bibinfo {volume}
  {41}},\ \bibinfo {pages} {042712} (\bibinfo {year} {2023})}\BibitemShut
  {NoStop}%
\bibitem [{\citenamefont {Bl{\"o}chl}(1994)}]{Blochl1994}%
  \BibitemOpen
  \bibfield  {author} {\bibinfo {author} {\bibfnamefont {P.~E.}\ \bibnamefont
  {Bl{\"o}chl}},\ }\href@noop {} {\bibfield  {journal} {\bibinfo  {journal}
  {Phys. Rev. B}\ }\textbf {\bibinfo {volume} {50}},\ \bibinfo {pages} {17953}
  (\bibinfo {year} {1994})}\BibitemShut {NoStop}%
\bibitem [{\citenamefont {Kresse}\ and\ \citenamefont
  {Hafner}(1993)}]{Kresse1993}%
  \BibitemOpen
  \bibfield  {author} {\bibinfo {author} {\bibfnamefont {G.}~\bibnamefont
  {Kresse}}\ and\ \bibinfo {author} {\bibfnamefont {J.}~\bibnamefont
  {Hafner}},\ }\href@noop {} {\bibfield  {journal} {\bibinfo  {journal} {Phys.
  Rev. B}\ }\textbf {\bibinfo {volume} {47}},\ \bibinfo {pages} {558} (\bibinfo
  {year} {1993})}\BibitemShut {NoStop}%
\bibitem [{\citenamefont {Kresse}\ and\ \citenamefont
  {Furthm{\"u}ller}(1996)}]{Kresse1996}%
  \BibitemOpen
  \bibfield  {author} {\bibinfo {author} {\bibfnamefont {G.}~\bibnamefont
  {Kresse}}\ and\ \bibinfo {author} {\bibfnamefont {J.}~\bibnamefont
  {Furthm{\"u}ller}},\ }\href@noop {} {\bibfield  {journal} {\bibinfo
  {journal} {Phys. Rev. B}\ }\textbf {\bibinfo {volume} {54}},\ \bibinfo
  {pages} {11169} (\bibinfo {year} {1996})}\BibitemShut {NoStop}%
\bibitem [{\citenamefont {Heyd}\ \emph {et~al.}(2003)\citenamefont {Heyd},
  \citenamefont {Scuseria},\ and\ \citenamefont {Ernzerhof}}]{Heyd2003}%
  \BibitemOpen
  \bibfield  {author} {\bibinfo {author} {\bibfnamefont {J.}~\bibnamefont
  {Heyd}}, \bibinfo {author} {\bibfnamefont {G.~E.}\ \bibnamefont {Scuseria}},
  \ and\ \bibinfo {author} {\bibfnamefont {M.}~\bibnamefont {Ernzerhof}},\
  }\href {\doibase 10.1063/1.1564060} {\bibfield  {journal} {\bibinfo
  {journal} {J. Chem. Phys.}\ }\textbf {\bibinfo {volume} {118}},\ \bibinfo
  {pages} {8207} (\bibinfo {year} {2003})}\BibitemShut {NoStop}%
\bibitem [{\citenamefont {Heyd}\ \emph {et~al.}(2006)\citenamefont {Heyd},
  \citenamefont {Scuseria},\ and\ \citenamefont {Ernzerhof}}]{Heyd2006}%
  \BibitemOpen
  \bibfield  {author} {\bibinfo {author} {\bibfnamefont {J.}~\bibnamefont
  {Heyd}}, \bibinfo {author} {\bibfnamefont {G.~E.}\ \bibnamefont {Scuseria}},
  \ and\ \bibinfo {author} {\bibfnamefont {M.}~\bibnamefont {Ernzerhof}},\
  }\href {\doibase 10.1063/1.2204597} {\bibfield  {journal} {\bibinfo
  {journal} {J. Chem. Phys.}\ }\textbf {\bibinfo {volume} {124}},\ \bibinfo
  {pages} {219906} (\bibinfo {year} {2006})}\BibitemShut {NoStop}%
\bibitem [{\citenamefont {Mu}\ \emph {et~al.}(2019)\citenamefont {Mu},
  \citenamefont {Peelaers},\ and\ \citenamefont {{Van de Walle}}}]{Mu2019}%
  \BibitemOpen
  \bibfield  {author} {\bibinfo {author} {\bibfnamefont {S.}~\bibnamefont
  {Mu}}, \bibinfo {author} {\bibfnamefont {H.}~\bibnamefont {Peelaers}}, \ and\
  \bibinfo {author} {\bibfnamefont {C.~G.}\ \bibnamefont {{Van de Walle}}},\
  }\href {\doibase 10.1063/1.5131755} {\bibfield  {journal} {\bibinfo
  {journal} {Appl. Phys. Lett.}\ }\textbf {\bibinfo {volume} {115}},\ \bibinfo
  {pages} {242103} (\bibinfo {year} {2019})}\BibitemShut {NoStop}%
\bibitem [{\citenamefont {Sabino}\ \emph {et~al.}(2014)\citenamefont {Sabino},
  \citenamefont {{de Oliveira}},\ and\ \citenamefont {Da~Silva}}]{Sabino2014}%
  \BibitemOpen
  \bibfield  {author} {\bibinfo {author} {\bibfnamefont {F.~P.}\ \bibnamefont
  {Sabino}}, \bibinfo {author} {\bibfnamefont {L.~N.}\ \bibnamefont {{de
  Oliveira}}}, \ and\ \bibinfo {author} {\bibfnamefont {J.~L.~F.}\ \bibnamefont
  {Da~Silva}},\ }\href {\doibase 10.1103/PhysRevB.90.155206} {\bibfield
  {journal} {\bibinfo  {journal} {Phys. Rev. B}\ }\textbf {\bibinfo {volume}
  {90}},\ \bibinfo {pages} {155206} (\bibinfo {year} {2014})}\BibitemShut
  {NoStop}%
\bibitem [{\citenamefont {{\AA}hman}\ \emph {et~al.}(1996)\citenamefont
  {{\AA}hman}, \citenamefont {Svensson},\ and\ \citenamefont
  {Albertsson}}]{Aahman1996}%
  \BibitemOpen
  \bibfield  {author} {\bibinfo {author} {\bibfnamefont {J.}~\bibnamefont
  {{\AA}hman}}, \bibinfo {author} {\bibfnamefont {G.}~\bibnamefont {Svensson}},
  \ and\ \bibinfo {author} {\bibfnamefont {J.}~\bibnamefont {Albertsson}},\
  }\href@noop {} {\bibfield  {journal} {\bibinfo  {journal} {Acta Crystallogr.
  C}\ }\textbf {\bibinfo {volume} {52}},\ \bibinfo {pages} {1336} (\bibinfo
  {year} {1996})}\BibitemShut {NoStop}%
\bibitem [{\citenamefont {Matsumoto}\ \emph
  {et~al.}(1974{\natexlab{b}})\citenamefont {Matsumoto}, \citenamefont {Aoki},
  \citenamefont {Kinoshita},\ and\ \citenamefont {Aono}}]{Matsumoto1974b}%
  \BibitemOpen
  \bibfield  {author} {\bibinfo {author} {\bibfnamefont {T.}~\bibnamefont
  {Matsumoto}}, \bibinfo {author} {\bibfnamefont {M.}~\bibnamefont {Aoki}},
  \bibinfo {author} {\bibfnamefont {A.}~\bibnamefont {Kinoshita}}, \ and\
  \bibinfo {author} {\bibfnamefont {T.}~\bibnamefont {Aono}},\ }\href@noop {}
  {\bibfield  {journal} {\bibinfo  {journal} {Jpn. J. Appl. Phys.}\ }\textbf
  {\bibinfo {volume} {13}},\ \bibinfo {pages} {1578} (\bibinfo {year}
  {1974}{\natexlab{b}})}\BibitemShut {NoStop}%
\bibitem [{\citenamefont {Haan}(1962)}]{Haan1962}%
  \BibitemOpen
  \bibfield  {author} {\bibinfo {author} {\bibfnamefont {Y.~D.}\ \bibnamefont
  {Haan}},\ }\href@noop {} {\bibfield  {journal} {\bibinfo  {journal} {Z.
  Krist.-Cryst. Mate.}\ }\textbf {\bibinfo {volume} {117}},\ \bibinfo {pages}
  {235} (\bibinfo {year} {1962})}\BibitemShut {NoStop}%
\bibitem [{\citenamefont {French}(1990)}]{French1990}%
  \BibitemOpen
  \bibfield  {author} {\bibinfo {author} {\bibfnamefont {R.~H.}\ \bibnamefont
  {French}},\ }\href@noop {} {\bibfield  {journal} {\bibinfo  {journal} {J. Am.
  Ceram. Soc.}\ }\textbf {\bibinfo {volume} {73}},\ \bibinfo {pages} {477}
  (\bibinfo {year} {1990})}\BibitemShut {NoStop}%
\bibitem [{\citenamefont {Marezio}(1966)}]{Marezio1966}%
  \BibitemOpen
  \bibfield  {author} {\bibinfo {author} {\bibfnamefont {M.}~\bibnamefont
  {Marezio}},\ }\href@noop {} {\bibfield  {journal} {\bibinfo  {journal} {Acta
  Crystallogr.}\ }\textbf {\bibinfo {volume} {20}},\ \bibinfo {pages} {723}
  (\bibinfo {year} {1966})}\BibitemShut {NoStop}%
\bibitem [{\citenamefont {Irmscher}\ \emph {et~al.}(2014)\citenamefont
  {Irmscher}, \citenamefont {Naumann}, \citenamefont {Pietsch}, \citenamefont
  {Galazka}, \citenamefont {Uecker}, \citenamefont {Schulz}, \citenamefont
  {Schewski}, \citenamefont {Albrecht},\ and\ \citenamefont
  {Fornari}}]{Irmscher2014}%
  \BibitemOpen
  \bibfield  {author} {\bibinfo {author} {\bibfnamefont {K.}~\bibnamefont
  {Irmscher}}, \bibinfo {author} {\bibfnamefont {M.}~\bibnamefont {Naumann}},
  \bibinfo {author} {\bibfnamefont {M.}~\bibnamefont {Pietsch}}, \bibinfo
  {author} {\bibfnamefont {Z.}~\bibnamefont {Galazka}}, \bibinfo {author}
  {\bibfnamefont {R.}~\bibnamefont {Uecker}}, \bibinfo {author} {\bibfnamefont
  {T.}~\bibnamefont {Schulz}}, \bibinfo {author} {\bibfnamefont
  {R.}~\bibnamefont {Schewski}}, \bibinfo {author} {\bibfnamefont
  {M.}~\bibnamefont {Albrecht}}, \ and\ \bibinfo {author} {\bibfnamefont
  {R.}~\bibnamefont {Fornari}},\ }\href@noop {} {\bibfield  {journal} {\bibinfo
   {journal} {Phys. Status Solidi A}\ }\textbf {\bibinfo {volume} {211}},\
  \bibinfo {pages} {54} (\bibinfo {year} {2014})}\BibitemShut {NoStop}%
\bibitem [{\citenamefont {Franciosi}\ and\ \citenamefont {{Van de
  Walle}}(1996)}]{Franciosi96}%
  \BibitemOpen
  \bibfield  {author} {\bibinfo {author} {\bibfnamefont {A.}~\bibnamefont
  {Franciosi}}\ and\ \bibinfo {author} {\bibfnamefont {C.~G.}\ \bibnamefont
  {{Van de Walle}}},\ }\href@noop {} {\bibfield  {journal} {\bibinfo  {journal}
  {Surface Science Reports}\ }\textbf {\bibinfo {volume} {25}},\ \bibinfo
  {pages} {1} (\bibinfo {year} {1996})}\BibitemShut {NoStop}%
\end{thebibliography}%

\end{document}